\documentclass[preprint,showpacs]{revtex4}
\usepackage{CJK}
\usepackage{amsmath}
\usepackage{graphicx}
\usepackage{braket}
\usepackage{color}

\begin{document}

\title{ Constraints on the R-parity violating couplings using the newest measurement of the decay $B_s^0 \rightarrow \mu^+ \mu^-$}
\author{Cheng Li}
\address{Institute of High Energy Physics and Theoretical Physics Center for Science Facilities,
Chinese Academy of Sciences, Beijing 100049, People¡¯s Republic of China}
\author{Xiangdong Gao}
\address{Institute of Physics, Academia Sinica, Nangang, Taipei 11529, Taiwan}
\author{Cai-Dian Lu}
\address{Institute of High Energy Physics and Theoretical Physics Center for Science Facilities,
Chinese Academy of Sciences, Beijing 100049, People¡¯s Republic of China}
\date{\today}

\begin{abstract}
Recently, the LHCb collaboration reported the first evidence for the decay $B_s^0 \rightarrow \mu^+ \mu^-$. A branching ratio of
$\mathcal{B}(B_s^0 \rightarrow \mu^+ \mu^-) = (3.2^{+1.5}_{-1.2}) \times 10^{-9}$ is given. Using the newest data, and together with the most precise predictions of the Standard Model contributions to the decay, we derive the constraints on the combinations of the R-parity violating parameters. Our results are several orders of magnitudes stronger than the  constraints in the previous literature. We also update the constraints on the relevant parameters using the upper limit of $\mathcal{B}(B_d^0 \rightarrow \mu^+ \mu^-)$.
\end{abstract}


\maketitle

\section{introduction}
The helicity suppressed rare decay process $B_s^0 \rightarrow \mu^+
\mu^-$ is induced by $Z$ boson mediated penguin diagram and box
diagram in the Standard Model (SM). This double suppression
mechanism makes the SM prediction for the process very small
\cite{Buras:2012ru}
\begin{equation}
\mathcal{B}(B_s^0 \rightarrow \mu^+ \mu^-) = (3.23 \pm 0.27) \times 10^{-9}.
\label{SMprediction}
\end{equation}
The fact that there are only leptons in the final states makes it a
golden channel for the discovery and/or constraining the new physics
model parameter space, since the new physics contributions can be
larger than the SM effects and there is the least hadronic
uncertainty.

The  minimal supersymmetric standard model with R-parity violation
(MSSM-RPV) is an extension of the  minimal supersymmetric standard
model (MSSM) by abandoning the discrete symmetry, the R-parity,
which is defined by $R_p = (-1)^{3B+L+2S}$, where B is the baryon
number, L is the lepton number, and S is the spin of the particle.
The most general R-parity violating term can be included in the MSSM
by introducing the following superpotential \cite{Weinberg:1981wj,
Sakai:1981pk}:
\begin{equation}
W_{\not{R_p}} = \mu_i H_u L_i + \frac{1}{2} \lambda_{ijk} L_i L_j E_k^c + \lambda_{ijk}^{\prime} L_i Q_j D_k^c + \frac{1}{2} \lambda_{ijk}^{\prime\prime} U_i^c D_j^c D_k^c,
\end{equation}
where additional factor of 1/2 is added because of the fact that the
first two indices of the couplings $\lambda_{ijk}$ and
$\lambda_{ijk}^{\prime\prime}$ are antisymmetric. It is easy to see
  from this superpotential that the $B_s^0 \rightarrow \mu^+
\mu^-$ decay can be induced at the tree level from the lepton number
violating terms $\lambda$ and $\lambda^{\prime}$.

Study of the MSSM-RPV have been performed in many rare decay
processes. The bounds on the relevant parameters in the MSSM-RPV
obtained from the decay $B_s^0 \rightarrow \mu^+ \mu^-$ were derived
in Ref. \cite{Jang:1997jy} and revised in the literature
\cite{Dreiner:2001kc, Saha:2002kt}. However, there were only upper bounds from experiments at
that time.

Recently, the LHCb collaboration reported the first measurement of
the branching ratio of $B_s^0 \rightarrow \mu^+ \mu^-$
\cite{:2012ct}
\begin{equation}
\mathcal{B}(B_s^0 \rightarrow \mu^+ \mu^-) = (3.2^{+1.5}_{-1.2}) \times 10^{-9}.\label{exp}
\end{equation}
This just lies on the central regions of the SM prediction in Eq.
(\ref{SMprediction}), which will put severe constraints on every new
physics  models. In this brief report, we will use the newest data
in Eq. (\ref{exp}) to constrain the relevant parameters in the
framework of the minimal supersymmetric standard model with R-parity
violation. Since the experimental measurement is quite close to the
SM prediction, we have to include the contributions of the standard
model together with the new physics contribution. Using the first
time measurement of the branching ratio of the decay process, we
give the most stringent constraints on the relevant parameters in
the MSSM-RPV. We also update the constraint on the sneutrino
exchange term and the squark exchange term from the newest
experimental upper limits \cite{:2012ct}:
  \begin{equation}
  \mathcal{B}(B_d^0 \rightarrow \mu^+ \mu^-) <9.4 \times 10^{-10} .\label{bd}
 \end{equation}

This brief report is organized as follows, in Sec. \ref{formalism}
we present the analytical expressions; and then we use these
equations to give the numerical results and discussions in Sec.
\ref{results}. We close this paper with a conclusion in Sec.
\ref{summary}.

\section{Analytical Expressions \label{formalism}}

In this section, we present the formalism for the calculation of the
branching ratio of the $B_s^0 \rightarrow \mu^+ \mu^-$ in both the
SM and the MSSM-RPV. The same formalism can also be applied to the
process $B_d^0 \rightarrow \mu^+ \mu^-$ with changing the
corresponding Cabibbo-Kobayashi-Maskawa   matrix elements. In
the SM, the effective Hamiltonian governing $B_s^0 \rightarrow \mu^+
\mu^-$ can be written as \cite{Buchalla:1995vs}
\begin{equation}
\mathcal{H}_{eff}^{\text{SM}} = -\frac{G_F}{\sqrt{2}}
\frac{\alpha}{2 \pi \sin^2 \Theta_W} V^{\ast}_{tb}V_{ts} \eta_Y
Y_0(x_t) (\bar{b}\gamma_\mu P_L s) (\bar{\mu}\gamma^\mu P_L \mu) +
h.c.,
\end{equation}
with $P_L=(1-\gamma_5)/2$. The function $Y_0$ is the famous
Inami-Lim function calculated from the electroweak penguin and box
diagrams \cite{Inami:1980fz}:
\begin{equation}
Y_0 (x) = \frac{x}{8} \left(\frac{4-x}{1-x} + \frac{3x}{(1-x)^2} \ln
x\right).
\end{equation}
 While $\eta_Y = 1.026 \pm 0.006$ includes higher order QCD corrections \cite{Buras:1998raa}.

The branching fraction of $B_s^0 \rightarrow \mu^+ \mu^-$ can be given by
\begin{equation}
\mathcal{B}(B_s^0 \rightarrow \mu^+ \mu^-) = \frac{\tau(B_s^0)
}{16\pi m_{B_s^0}} |\mathcal{H}_{eff}^{\text{SM}}|^2   \sqrt{1 -
\frac{4m_{\mu}^2}{m^2_{B_s^0}}}, \label{decayrate}
\end{equation}
where $\tau(B_s^0)$ is the lifetime of the $B_s^0$ meson.

\begin{figure}
\includegraphics[scale=0.65]{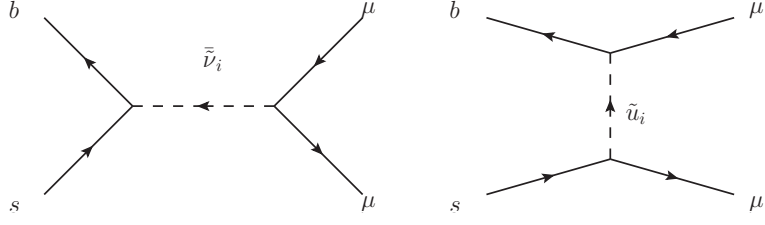}
\caption{Feynman diagrams contributing to $B_s^0 \rightarrow \mu^+
\mu^-$ in the minimal supersymmetric standard model with R-parity
violation.} \label{feynman}
\end{figure}

In the MSSM-RPV, the relevant effective Hamiltonian can be obtained
by matching the amplitudes in full theory as shown in Fig.
\ref{feynman} onto the effective four fermion operators. For the
left diagram of  Fig. \ref{feynman}, the effective Hamiltonian from
the exchange of sneutrino can be written as
\begin{equation}
\mathcal{H}_{eff}^{\text{RPV}} = -\frac{A}{2} (\bar{\mu}P_R \mu) (\bar{b}P_L s),
\end{equation}
where $P_R=(1+\gamma_5)/2$ and
\begin{equation}
A = \sum_i
\frac{\lambda^{\ast}_{i22}\lambda^{\prime}_{i23}}{m_{\tilde{\nu}_i}^2}.
\end{equation}
For this new kind of contribution, the interference of the
$(\bar{\mu}P_R \mu) $ density with the standard model
$(\bar{\mu}\gamma^\mu P_L \mu)$ current leads to zero, so we can
directly separate the SM and the MSSM-RPV contributions to the
branching ratio as
\begin{equation}
\mathcal{B}(B_s^0 \rightarrow \mu^+ \mu^-) = \mathcal{B}^{SM} + \mathcal{B}^{RPV}_A,
\end{equation}
where we will use Eq. (\ref{SMprediction}) as input value of
$\mathcal{B}^{SM}$ in our numerical calculations, while
\begin{equation}
\mathcal{B}^{RPV}_A = |A|^2 \frac{\tau(B_s^0)}{16\pi} f_{B_s^0}^2
m_{B_s^0}^3 \left(1 - \frac{2m_{\mu}^2}{m_{B_s^0}^2}\right)
\sqrt{1-\frac{4m_{\mu}^2}{m_{B_s^0}^2}}.
\end{equation}

For the up-squark contribution shown in the right diagram of Fig.
\ref{feynman}, the effective Hamiltonian is
\begin{equation}
\mathcal{H}_{eff}^{\text{RPV}} = + \frac{B}{8} (\bar{b}\gamma_\mu
P_R s) (\bar{\mu}\gamma^\mu P_L \mu),
\end{equation}
where
\begin{equation}
B = \sum_i
\frac{\lambda^{\prime\ast}_{2i2}\lambda^{\prime}_{2i3}}{m_{\tilde{u}_i}^2}.
\end{equation}
For this scalar quark contribution has the same structure of current
as the standard model case, and the SM prediction lies in the
central values of the experimental data, we can reasonably assume
that the interference term of the SM and the MSSM-RPV is greatly
larger than the pure MSSM-RPV contributions, so we will
approximately write the total contributions as
\begin{equation}
\mathcal{B}(B_s^0 \rightarrow \mu^+ \mu^-) = \mathcal{B}^{SM} + \mathcal{B}^{int}_B,
\end{equation}
where we also use the numerical value for $\mathcal{B}^{SM}$ from
Eq. (\ref{SMprediction}); while the interference term can be written
as:
\begin{equation}
\mathcal{B}^{int}_B = B \frac{\tau(B_s^0) G_F
\alpha}{16\sqrt{2}\pi^2\sin^2 \Theta_W} V_{tb}V_{ts}^{\ast}
f_{B_s^0}^2 m_{B_s^0} m_{\mu}^2 \eta_Y
Y_0(x_t)
\sqrt{1-\frac{4m_{\mu}^2}{m_{B_s^0}^2}}.
\end{equation}

The $B_s^0$ meson decay constant $f_{B_s^0}$ shown in the above
equations arises from the calculation of the hadronic matrix
element, which is defined as
\begin{equation}
\bra{0}\bar{b}\gamma^{\mu}\gamma_5s\ket{B_s^0} = i f_{B_s^0} p_{B_s^0}^{\mu}.
\end{equation}
The hadronic matrix elements for the pseudo-scalar density can be
derived from the equation of motion under the assumption that $m_b
\simeq m_{B_s^0}$.

\section{Numerical results \label{results}}

In this section we present our numerical results. Following Ref.
\cite{Buras:2012ru}, the SM parameters are taken as $G_F = 1.16638
\times 10^{-5} \text{GeV}^{-2}$, $\alpha =1/127.937$, $m_W =
80.385$GeV \cite{Beringer:1900zz}, $m_t = 173.2$GeV
\cite{Aaltonen:2012ra}, $m_{\mu} = 105.6584$MeV,
$|V_{tb}^{\ast}V_{ts}| = 0.0405$ and $|V_{tb}^{\ast}V_{td}| =
0.0087$. The relevant parameters of neutral $B$ mesons are collected
in table \ref{parameter}. Uncertainties of these parameters are not
considered in the numerical calculation, since the uncertainties
induced by these parameters on the relevant constraints on the
couplings of the MSSM-RPV are far beyond the scope of the
experimental data.

\begin{table}
\caption{Input parameters for $B_s^0$ and $B_d^0$ mesons used in
numerical calculations. Uncertainties of these parameters are not
considered for the reason that the uncertainties induced by these
parameters on the relevant constraints on the couplings of the
MSSM-RPV are far beyond the scope of the experimental data.}
  \begin{tabular}{|c|c|c|c|}
    \hline & $\tau_B$(ps) & $f_B$(MeV) & $m_B$(GeV) \\
    \hline $B_s^0 $ & 1.466 & 227 & 5.36677 \\
    \hline $B_d^0 $ & 1.519 & 190 & 5.27958 \\
    \hline
  \end{tabular}
  \label{parameter}
\end{table}

As described in the last section, the contributions of the SM are
taken as input, and then we calculate the total effects from the sum
of the SM and the scalar neutrino contribution or the sum of SM and
scalar quark contribution of the MSSM-RPV. By demanding the total
contributions do not exceed the experimental upper and lower bounds,
we obtain the following constraints on the relevant combinations of
the parameters in the MSSM-RPV, respectively,
\begin{eqnarray}
 & |\sum_i \frac{\lambda^{\ast}_{i22}\lambda^{\prime}_{i23}}{m_{\tilde{\nu}_i}^2}|  & < 6.52\times10^{-11},
\nonumber\\
 -2.29\times10^{-9} <&\sum_i \frac{\lambda^{\prime\ast}_{2i1}\lambda^{\prime}_{2i3}}{m_{\tilde{u}_i}^2}&< 2.87\times10^{-9}.
\end{eqnarray}
The scalar neutrino coupling suffers roughly 2 orders of stronger
constraints than the scalar quark coupling since there
is no helicity suppression in the sneutrino contributions. Due to the more stringent experimental
limit, our results are several orders of magnitude stronger than
previous results in the literature \cite{Dreiner:2001kc,
Saha:2002kt}. If we further assume the mass of the sparticles to be
several hundreds GeV, a roughly estimate shows that the above
combinations of the R-parity violating couplings are around
$10^{-6}$ and $10^{-4}$, respectively, which means that the magnitudes of
the couplings are not too far away from unity.

We also give the constraints from the newest experimental upper
limits on the $\mathcal{B}(B_d^0 \rightarrow \mu^+ \mu^-)$ shown in
Eq. (\ref{bd}). The corresponding constraints on the sneutrino
exchange term and the squark exchange term are given below
\begin{eqnarray}
|\sum_i \frac{\lambda^{\ast}_{i22}\lambda^{\prime}_{i13}}{m_{\tilde{\nu}_i}^2}| & < 7.85\times10^{-11},
\nonumber\\
\sum_i \frac{\lambda^{\prime\ast}_{2i1}\lambda^{\prime}_{2i3}}{m_{\tilde{u}_i}^2} & < 1.17\times10^{-8}.
\end{eqnarray}

\section{Summary}\label{summary}

In conclusion, using the newest experimental data, we have
calculated the contributions to the $B_{s(d)}^0 \rightarrow \mu^+
\mu^-$ in the framework of the MSSM-RPV. We gave the constraints on
the relevant combinations of the parameters in the MSSM-RPV as
\begin{eqnarray}
 & |\sum_i \frac{\lambda^{\ast}_{i22}\lambda^{\prime}_{i23}}{m_{\tilde{\nu}_i}^2}|  & < 6.52\times10^{-11},
\nonumber\\
 -2.29\times10^{-9} <&\sum_i \frac{\lambda^{\prime\ast}_{2i1}\lambda^{\prime}_{2i3}}{m_{\tilde{u}_i}^2}&< 2.87\times10^{-9};
\end{eqnarray}
\begin{eqnarray}
|\sum_i \frac{\lambda^{\ast}_{i22}\lambda^{\prime}_{i13}}{m_{\tilde{\nu}_i}^2}| & < 7.85\times10^{-11},
\nonumber\\
\sum_i \frac{\lambda^{\prime\ast}_{2i1}\lambda^{\prime}_{2i3}}{m_{\tilde{u}_i}^2} & < 1.17\times10^{-8}.
\end{eqnarray}
Our results are several orders of magnitude stronger than the
previous results in the literature.

\section*{Acknowledgements}
The work of C. L and C. D. L. is partly supported by the National Science Foundation of China under the Grant No.11075168, 11228512
and 11235005. The work of X. D. G. is partly supported by the National Science Council of Taiwan under Grant No 101-2811-M-001-060-.

\bibliography{references}

\end{document}